\documentclass{nature}
\usepackage{graphicx,psfrag}
\usepackage{mathrsfs}
\usepackage{mhchem}
\usepackage{amsmath,amsfonts,amssymb}
\usepackage{multirow}
\usepackage{comment}
\usepackage{units}
\usepackage{enumerate}
\usepackage[normalem]{ulem} 
\usepackage{longtable}
\usepackage[rgb]{xcolor}
\usepackage{gensymb}
\usepackage{adjustbox}
\usepackage[none]{hyphenat}
\usepackage{multibib}
\newcites{main}{Main References}
\newcites{methods}{Methods References}
\usepackage{caption}
\usepackage{url}
\newcites{New}{References}
\usepackage{lineno}
\usepackage{longtable}

\newcommand{\mn}{{Mon. Not. R. Astron. Soc.}}
\newcommand{\mnras}{\mn}
\newcommand{\aj}{{"Astron. J."}}
\newcommand{\apj}{{Astrophys. J.}}
\newcommand{\apjl}{{Astrophys. J. Lett.}}
\newcommand{\apjs}{{Astrophys. J. Supp.}}

\newcommand{\aap}{{Astron. Astrophys.}}
\newcommand{\nat}{{Nature}}

\newcommand{\pasp}{{Pub. Ast. Soc. Pac.}}

\title{A gravitational wave detectable candidate Type Ia supernova progenitor}

\author{Emma T. Chickles $^{1,2}$ $^{*}$, Kevin B. Burdge$^{1,2}$, Joheen Chakraborty$^{1,2}$, Vik S. Dhillon$^{3,4}$, Paul Draghis$^{1,2}$, James Munday$^{5}$, Saul A. Rappaport$^{1,2}$, John Tonry$^{6}$, Evan Bauer$^{7}$, Alex J. Brown$^{8}$, Noel Castro$^{5}$, Deepto Chakrabarty$^{1,2}$, Martin Dyer$^{3}$, Kareem El-Badry$^{9}$, Anna Frebel$^{1,2}$, Gabor Furesz$^{1,2}$, James Garbutt$^{3}$, Matthew J. Green$^{10}$, Aaron Householder$^{11,2}$, Scott A. Hughes$^{1,2}$, Daniel Jarvis$^{3}$, Erin Kara$^{1,2}$, Mark R. Kennedy$^{12}$, Paul Kerry$^{3}$, Stuart P Littlefair$^{3}$, James McCormac$^{5}$, Geoffrey Mo$^{1,2}$, Mason Ng$^{13,14,1,2}$, Steven Parsons$^{3}$, Ingrid Pelisoli$^{5}$, Eleanor Pike$^{3}$, Thomas A. Prince$^{9}$, George R. Ricker$^{1,2}$, Jan van Roestel$^{15}$, David Sahman$^{3}$, Ken J. Shen$^{16}$, Robert A. Simcoe$^{1,2}$, Pier-Emmanuel Tremblay$^{5}$, Andrew Vanderburg$^{1,2}$, Tin Long Sunny Wong$^{17}$}

\begin{document}

\maketitle

\begin{affiliations}
 \item Department of Physics, Massachusetts Institute of Technology, Cambridge, MA 02139, USA
 \item Kavli Institute for Astrophysics and Space Research, Massachusetts Institute of Technology, Cambridge, MA 02139, USA
 \item Astrophysics Research Cluster, School of Mathematical and Physical Sciences, University of Sheffield, Sheffield S3 7RH, UK
 \item Instituto de Astrof\'{i}sica de Canarias, E-38205 La Laguna, Tenerife, Spain
 \item Department of Physics, University of Warwick, Coventry CV4 7AL, UK
 \item Institute for Astronomy, University of Hawaii, 2680 Woodlawn Drive, Honolulu, HI 96822-1897, USA
 \item Center for Astrophysics, Harvard \& Smithsonian, 60 Garden Street, Cambridge, MA 02138
 \item Hamburger Sternwarte, University of Hamburg, Gojenbergsweg 112, 21029 Hamburg, Germany
 \item Division of Physics, Mathematics and Astronomy, California Institute of Technology, Pasadena, CA, USA
 \item Max-Planck-Institut f\"{u}r Astronomie, K\"{o}nigstuhl 17, D-69117 Heidelberg, Germany
 \item Department of Earth, Atmospheric and Planetary Sciences, Massachusetts Institute of Technology, Cambridge, MA 02139, USA
 \item Department of Physics, University College Cork, Cork, Ireland
 \item Department of Physics, McGill University, 3600 rue University, Montr\'{e}al, QC H3A 2T8, Canada
 \item Trottier Space Institute, McGill University, 3550 rue University, Montr\'{e}al, QC H3A 2A7, Canada
 \item Anton Pannekoek Institute for Astronomy, University of Amsterdam, 1090 GE Amsterdam, The Netherlands
 \item Department of Astronomy and Theoretical Astrophysics Center, University of California, Berkeley, CA 94720, USA
 \item Department of Physics, University of California, Santa Barbara, CA 93106, USA

 \end{affiliations}

\begin{abstract}

Type Ia supernovae, critical for studying cosmic expansion\citemain{1998AJ....116.1009R}$^,$\citemain{1998ApJ...507...46S}, arise from thermonuclear explosions of white dwarfs\citemain{1960ApJ...132..565H}, but their precise progenitor pathways remain unclear. Growing evidence supports the ``double-degenerate'' scenario, where two white dwarfs interact\citemain{1984ApJS...54..335I}$^,$\citemain{1984ApJ...277..355W}. The absence of other companion types capable of explaining the observed Ia rate\citemain{1984ApJ...277..355W}, along with observations of hyper-velocity white dwarfs interpreted as surviving companions of such systems\citemain{2018ApJ...865...15S}$^,$\citemain{2023OJAp....6E..28E} provide compelling evidence in favor of this scenario. Upcoming millihertz gravitational wave observatories like the Laser Interferometer Space Antenna (LISA)\citemain{2017arXiv170200786A} are expected to detect thousands of double-degenerate systems, though the most compact known candidate Ia progenitors produce only marginally detectable gravitational wave signals. Here, we report observations of ATLAS J1138-5139, a binary white dwarf system with an orbital period of just 28 minutes. Our analysis reveals a 1 solar mass carbon-oxygen white dwarf accreting from a helium-core white dwarf. Given its mass, the accreting carbon-oxygen white dwarf is poised to trigger a typical-luminosity Type Ia supernova within a few million years, or to evolve into a stably mass-transferring AM CVn system. ATLAS J1138-5139 provides a rare opportunity to calibrate binary evolution models by directly comparing observed orbital parameters and mass transfer rates closer to merger than any previously identified candidate Type Ia progenitor. Its compact orbit ensures detectability by LISA, demonstrating the potential of millihertz gravitational wave observatories to reveal a population of Type Ia progenitors on a Galactic scale, paving the way for multi-messenger studies offering insights into the origins of these cosmologically significant explosions.
\end{abstract}

The Asteroid Terrestrial-impact Last Alert System (ATLAS)\citemain{2018PASP..130f4505T} is a synoptic survey that images the entire sky every night. Using ATLAS, we searched for periodic brightness variations of 1.3 million white dwarf candidates\citemain{2021yCat..75083877G} as part of an ongoing campaign to identify binary systems in ultracompact orbits (Methods). Among them, ATLAS J1138-5139 stood out with a large-amplitude sinusoidal light curve, suggesting ``ellipsoidal'' variations caused by the tidal distortion of one star by an unseen companion. A simultaneous analysis of the Transiting Exoplanet Survey Satellite (TESS)\citemain{2015JATIS...1a4003R} data confirmed an orbital period of just 27.68 minutes; in order to fit into such a short orbit, both stars must be dense objects---white dwarfs.

Follow-up with ULTRACAM\citemain{2002ASPC..261..672D}, a high-speed imaging camera on the 3.5 m New Technology Telescope at the La Silla Observatory, revealed signatures of accretion, including a pronounced ``O’Connell effect'': the presence of unequal maxima (near orbital phases 0.25 and 0.75, Figure 1). Since the white dwarfs are in quadrature at time of maximum brightness, a detached binary system should exhibit the same brightness half an orbital period later\citemain{2009SASS...28..107W}. However, the peak fluxes of the alternating maxima in ATLAS J1138-5139 differ by 15\% in the $u_s$  filter, and by 7\% in the $r_s$ filter. These large amplitudes are inconsistent with the signal from relativistic Doppler beaming\citemain{1987A&A...183L..21S}, which could at most contribute $\sim1\%$ of the flux given the radial velocity semi-amplitude of the luminous component in the binary system (Methods). The strong wavelength dependence of the amplitude suggests that the source responsible for the O’Connell effect emits mostly at shorter wavelengths, indicating that it is originating from a region of significantly higher temperature than the $9350\,\rm K$ donor star. This ``hot spot'' is likely formed where the mass-transfer stream impacts the outer edge of the accretion disk surrounding the primary white dwarf. As a result, the hot spot is hidden from view by the accretion disk and/or the primary white dwarf during some quadratures (phase 0.25), but not at others (phase 0.75), resulting in the differential maxima.

In addition, the binary system's orientation with respect to our line of sight ($i>76^{\circ}$, Methods) enables us to observe eclipses causing periodic dips in brightness from obscuring the primary (at phase 0 in Figure 1) and secondary (at phase 0.5) white dwarfs. The primary eclipse is asymmetric: the ingress (the decline in brightness) takes less time than the egress (the increase in brightness). This asymmetry supports a geometry where the accretion disk and hot spot contribute to the observed brightness. The disk and the primary white dwarf are first eclipsed, followed by the ingress of the deflected hot spot. After the white dwarf egress (the sudden brightening around phase 0.04), the hot spot gradually becomes fully visible until phase 0.1, consistent with the expected behavior of an accretion flow. This timing of events in the light curve reinforces the interpretation that mass transfer is occurring in the system.

The presence of mass transfer indicates the lower mass secondary fills its Roche-lobe, donating mass to the accretor. A key result from binary star physics\citemain{1983ApJ...268..368E} is that the mean density of all Roche-filling objects at a given orbital period can be approximated by the following relation:
\[\bar{\rho}_{\text{donor}}\approx 0.185\,\text{g cm}^{-3}\times\left(\frac{P_{\text{orb}}}{\text{day}}\right)^{-2}\]
where $\bar{\rho}_{\text{donor}}$ is the mean density of the Roche-filling donor. This expression is accurate to within $6\%$ for mass ratios $0.01<q<1$, a range which encompasses all plausible mass ratios in this system.

Additionally, \emph{Gaia}'s geometric parallax measurement of ATLAS J1138-5139 allows us to measure the distance to the source, enabling us to determine the radius of the donor from its temperature and apparent luminosity. Knowing both the radius and the mean density allows us to infer the donor's mass. By combining the inferred mass and orbital inclination constraints with the precisely measured radial velocity semi-amplitude of 687.4 $\pm$ 3.8 km s$^{-1}$, obtained from spectroscopic observations using the Magellan Echellette (MagE) Spectrograph (Methods), we robustly determine the mass of the accretor white dwarf to be $1.02\pm0.04\,M_\odot$ (as shown in Figure 3), based solely on the Roche geometry and Kepler's laws. 

This system is the only ultracompact (that is, with an orbital period under an hour) binary known to host such a massive white dwarf. While massive white dwarfs have been identified in longer-period binaries, such as the 0.97 $M_\odot$ white dwarf in KPD 1930+2752 (2.283 hours)\citemain{2000MNRAS.317L..41M}$^,$\citemain{2007A&A...464..299G}, the $>1\,M_\odot$ white dwarf in V458 Vulpeculae (1.635 hours)\citemain{2010MNRAS.407L..21R}, and the $1.01\pm0.15\,M_\odot$ white dwarf in HD 265435 (99 minutes)\citemain{2021NatAs...5.1052P}, these systems are not expected to begin mass transfer for tens of millions of years, as gravitational wave emission must first significantly shrink the orbit. The scarcity of massive white dwarfs in ultracompact orbits suggests that ATLAS J1138-5139 may represent a short-lived evolutionary stage, rapidly evolving due to gravitational radiation.

The aforementioned longer-period systems have been predicted as Type Ia supernova (SN Ia) progenitors through the ``super-Chandrasekhar" channel. In this scenario, a merger between the two components would result in a combined mass exceeding the Chandrasekhar limit of $1.4\,M_\odot$, potentially triggering a supernova\citemain{2006Natur.443..308H}. However, the predicted number of super-Chandrasekhar white dwarf binaries is likely insufficient to explain the observed rate of SNe Ia\citemain{2012ApJ...749L..11B}. Furthermore, there is ongoing debate about the fate of these systems, with some studies suggesting that super-Chandrasekhar systems might collapse into a single compact object rather than fully detonating in a supernova\citemain{2004ApJ...615..444S}. Recent evidence supports a different progenitor pathway, in which the detonation of accreted helium triggers a secondary detonation in the sub-Chandrasekhar mass accretor, leading to a Type Ia supernova and ejecting the donor as a hypervelocity star\citemain{2018ApJ...865...15S}$^,$ \citemain{2023OJAp....6E..28E}. This scenario is particularly relevant for ATLAS J1138-5139, whose donor star resembles D6-2 --a hypervelocity star interpreted as a former helium white dwarf donor ejected by a double detonation\citemain{2021ApJ...923L..34B}.

The absence of helium features in the spectrum of ATLAS J1138-5139 implies that the donor star's hydrogen shell is sufficiently thick (approximately $10^{-4}-10^{-3}\,M_\odot$) to obscure underlying helium. Over the next few million years, the hydrogen layer will be gradually stripped away, eventually exposing helium-rich material. At a low accretion rate ($\lesssim 10^{-8}\,M_\odot/$yr)\citemain{2011ApJ...734...38W}$^,$\citemain{2016A&A...589A..43N}, the accumulated helium shell on the accretor white dwarf would remain sufficiently cool to allow stable mass transfer without ignition on the surface of the accretor, evolving into a stably mass-transferring AM Canum Venaticorum-type system with expanding orbital separation. 

However, if the accretion rate exceeds $\gtrsim10^{-8}\,M_\odot$/yr, the helium shell could reach temperatures that trigger a detonation on the surface of the accretor white dwarf. While such a helium nova could result in the accretor expanding and overflowing its Roche lobe, resulting in the binary merging as an R Coronae Borealis star\citemain{2015ApJ...805L...6S}, models of such detonations predict that at this white dwarf mass, the shock wave propagating into the underlying carbon-oxygen core will initiate runaway thermonuclear fusion resulting in a complete explosion of the white dwarf---a type Ia supernova. Binary population studies\citemain{2009ApJ...699.2026R} predict there exists a sufficient number of white dwarf binaries with sub-Chandrasekhar primary white dwarfs to explain the observed rate of SNe Ia. 

The derived masses along with the precisely determined orbital period allow us to constrain the merger time due to gravitational wave emission, which is found to be $\sim5.5$ million years. The characteristic gravitational wave strain of the system places it well above the detection limit of LISA (Figure 4), with a predicted signal-to-noise ratio of 6.51 by the end of LISA's 4-year mission. This is significantly higher than previous SNa Ia candidates in the literature, which all have SNR$\lesssim 1.5$\citemain{2000MNRAS.317L..41M,2007A&A...464..299G,2010MNRAS.407L..21R,2021NatAs...5.1052P}. ATLAS J1138-5139 is thus the very first demonstration of LISA's capability to blindly detect candidate SNe Ia progenitor systems through gravitational wave signals alone. Electromagnetic follow-up observations, such as those outlined in this study, will be critical in studying LISA's detections and constraining the rates and characteristics of white dwarf binaries that will produce SNe Ia, as a gravitational wave detection alone is insufficient to understand whether these sources are likely to result in a SNe Ia. This progenitor population will help quantify the relative contribution of the double-degenerate channel to the overall SNe Ia rate, showcasing the power of multi-messenger astronomy to answer longstanding questions about the nature of SNe Ia.

\begin{figure}
\includegraphics[width=\textwidth]{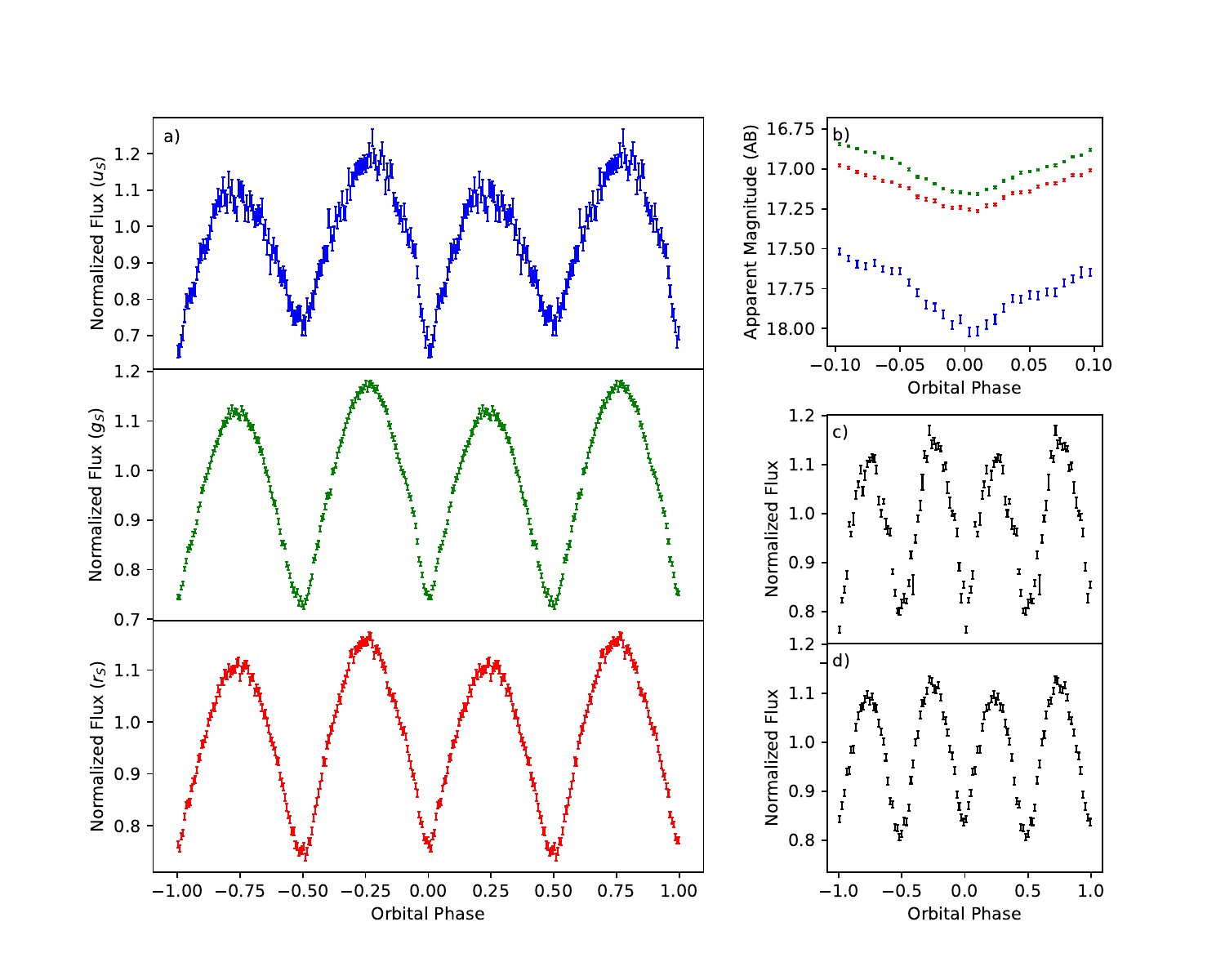}
\caption{\textbf{Light curve of ATLAS J1138-5139.} \textbf{a)} The binned $u_sg_sr_s$ ULTRACAM light curve of ATLAS J1138-5139, phase-folded on the 27.68-min orbital period. The light curve exhibits sinusoidal variations, with maxima at phases $0.25$ and $0.75$, due to an ellipsoidally deformed secondary star. At phase $0.75$, a heated accretion feature contributes additional flux, resulting in the appearance of unequal maxima. \textbf{b)} The asymmetric primary eclipse in the \emph{ugr} ULTRACAM light curve, indicating the presence of a luminous hot spot on the outer edge of an accretion disk surrounding the primary star. \textbf{c)} The binned and phase-folded light curve of the object from ATLAS. We were able to discover the object because of its periodic behavior. \textbf{d)} The binned and phase-folded TESS light curve.}
\label{fig:LC}
\end{figure}

\begin{figure}
\includegraphics[width=\textwidth]{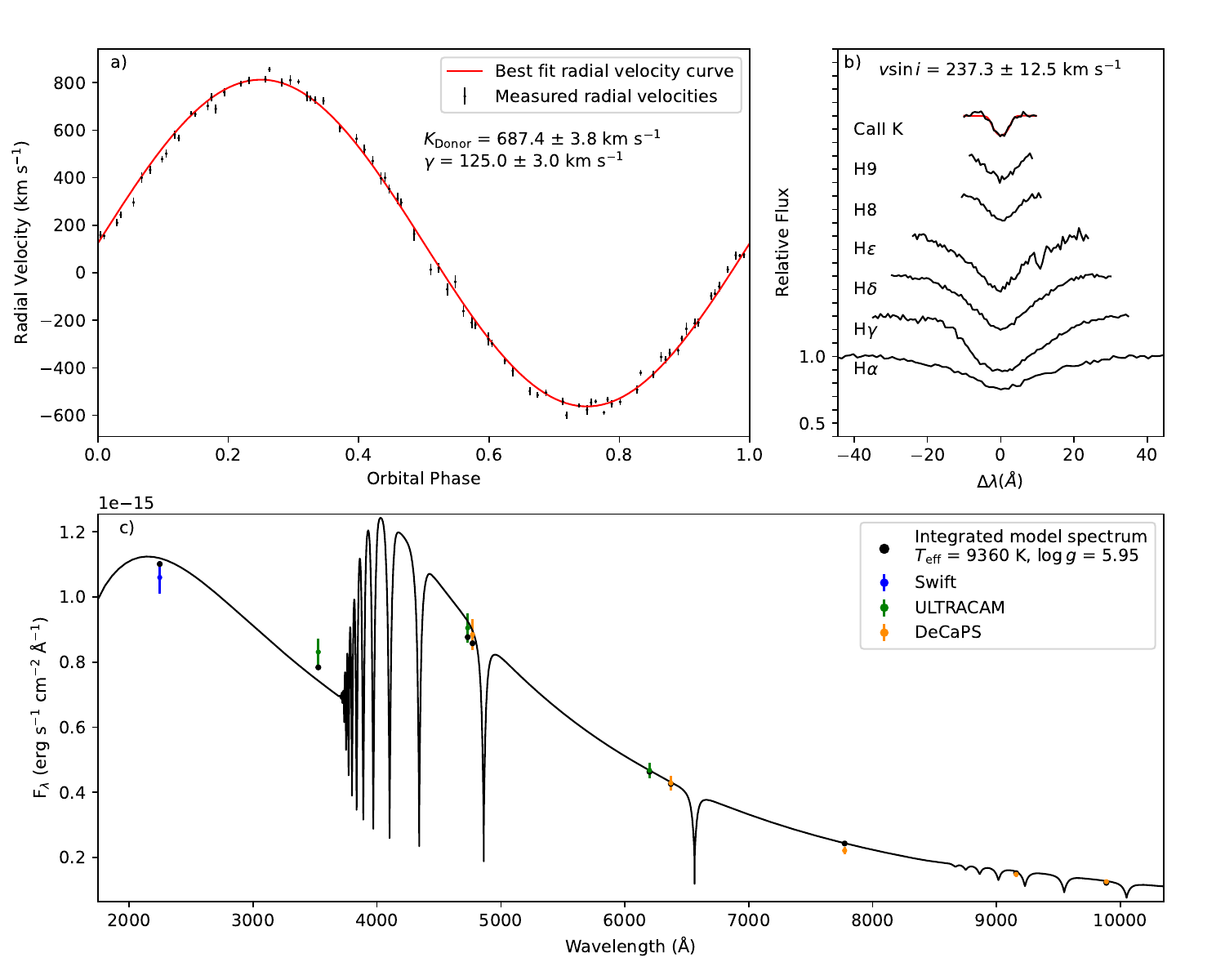}
\caption{\textbf{Optical spectroscopy and broadband photometry of ATLAS J1138-5139. a)} A sinusoidal fit to the measured radial velocities of the donor star in ATLAS J1138-5139, with a best-fit velocity semi-amplitude of $K_2=687.4\pm3.8\rm \, km\, s^{-1}$ and systemic velocity of $\gamma=125.0\pm3.0\rm \, km\, s^{-1}$. \textbf{b)} Optical spectrum of ATLAS J1138-5139 coadded in the rest frame of the donor star. Overlaid on the Ca II K line is a best-fit model of a Calcium-polluted white dwarf atmosphere, from which we inferred the rotational velocity $v\sin i=237.3\pm12.5\rm \, km\, s^{-1}$. \textbf{c)} The spectral energy distribution (SED) of ATLAS J1138-5139, with a best-fit model of a hydrogen-dominated white dwarf atmosphere (black solid line). The blue error bar shows the flux in the Swift UVM2 filter, the green error bars show the fluxes in the ULTRACAM filters, and the orange error bars show the fluxes in the DeCaPS filters. The black points show the corresponding synthetic fluxes in the different filters.}

\label{fig:Spectrum}
\end{figure}

\begin{figure}
\includegraphics[width=\textwidth]{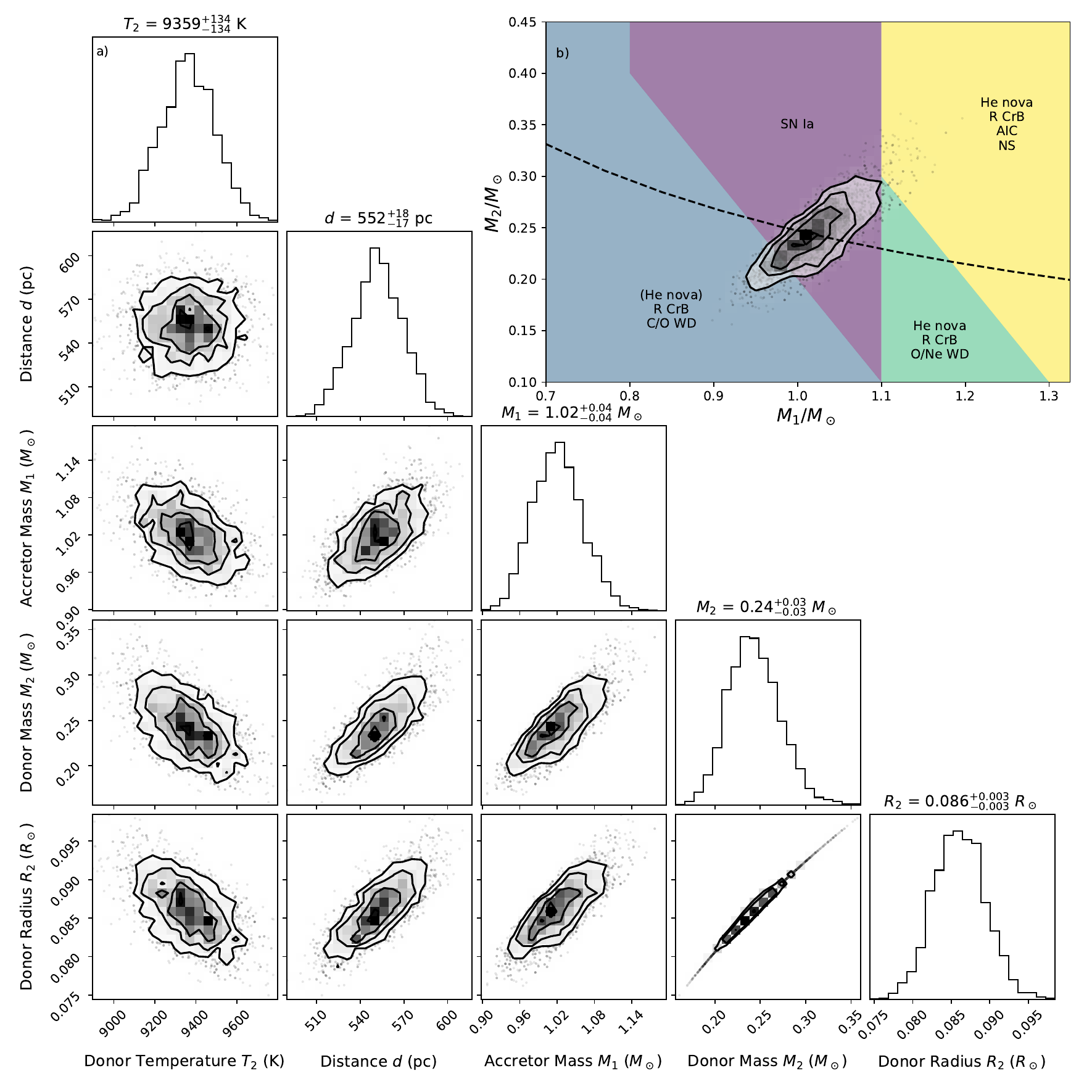}
\caption{\textbf{Constraints on the component masses, donor temperature and radius, and distance.} \textbf{a)} Corner plots for the fitting procedure. We simultaneously fit the SED data with the \emph{Gaia} astrometric constraint, and the spectroscopic radial velocity constraint, but assuming the same density since the donor is filling its Roche lobe. \textbf{b)} Outcomes of interacting double white dwarf binaries from Shen 2015\cite{2015ApJ...805L...6S}, with the posterior distributions of the component masses of ATLAS J1138-5139 and a dashed line of constant chirp mass.}
\label{fig:corner}
\end{figure}

\begin{figure}
\includegraphics[width=6.5in]{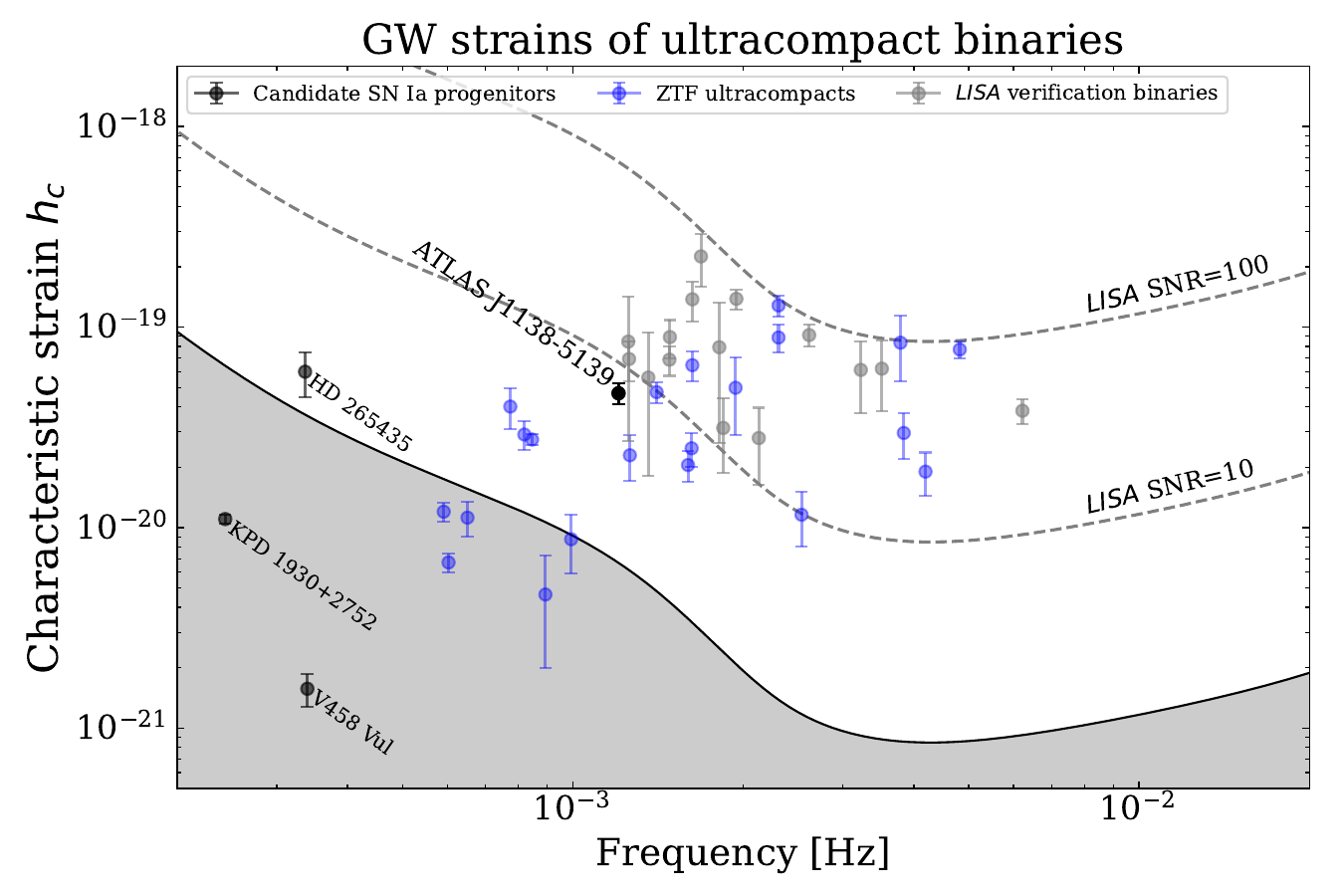}
\linespread{1.3}\selectfont{}
\caption{4-year LISA sensitivity curve detectability showing that ATLAS J1138-5139 will be blindly detectable with SNR=6.51. We also show the characteristic strains of three other SN Ia progenitor candidates from the literature\cite{2000MNRAS.317L..41M,2007A&A...464..299G,2010MNRAS.407L..21R,2021NatAs...5.1052P}, as well as ZTF-discovered ultracompact binaries\cite{2020ApJ...905...32B,2020ApJ...905L...7B,2024arXiv241112796C} and \textit{LISA} verification binaries\cite{2024ApJ...963..100K}.}
\end{figure}

\begin{table}
\renewcommand{\thetable}{\arabic{table}}
\centering
 \label{tab:Parameters}
\begin{tabular}{cc}

\hline
\hline
    $\rm Right\,\,ascension$ &$11^\mathrm{h}\,\,38^\mathrm{m}\,\,10.91^\mathrm{s}$ \\
   \hline
   $\rm Declination$ &$-51^\circ\,\,39'\,\,49.2''$ \\
   \hline
   $\rm Proper\,\,motion\,\,in\,\,right\,\,ascension$ &$-29.36\pm0.05\,\mathrm{mas\,yr^{-1}}$\\
   \hline
   $\rm Proper\,\,motion\,\,in\,\,declination$ &$3.65\pm0.05\,\mathrm{mas\,yr^{-1}}$\\
   \hline
   $\mathrm{Parallax}\,\,(\sigma)$ &$1.79\pm0.06\,\mathrm{mas}$\\
   \hline
   $\mathrm{Distance}\,\,(D)$ &$553^{+16}_{-18}\,\mathrm{pc}$\\
   \hline
   $\mathrm{Systemic\,\,velocity}\,\,(\gamma)$ &$125.0\pm3.0\,\mathrm{km\,s^{-1}}$\\   
   \hline
   \hline
   $\mathrm{Orbital\,\,period}\,\,(P_{\mathrm{b}})$ & $1660.92028(33)\,  \mathrm{s}$  \\
   \hline
   $\mathrm{Time\,\,of\,\,superior\,\,conjunction}\,\,(T_{0})$ & $60297.2822351\pm0.0000020\,   \mathrm{BMJD_{TDB}}$  \\
   \hline
   $\mathrm{Radial\,\,velocity\,\,of\,\,donor}\,\,(K_{\mathrm{Donor}})$ & $687.4\pm3.8\,\mathrm{km\,s^{-1}}$  \\
   \hline
   $\mathrm{Projected\,\,rotational\,\,velocity\,\,of\,\,donor}\,\,(v\mathrm{sin}i_{\mathrm{Donor}})$ & $237.3\pm12.5\,\mathrm{km\,s^{-1}}$  \\
   \hline
   $\mathrm{Orbital\,\,inclination}\,\,(i)$ & $>76^{\circ}\,\mathrm{(Eclipses)}$, $\sim88.6^{\circ}\mathrm{\,\, (Lightcurve\,\,model)}$  \\
    \hline
   $\mathrm{Semi\,\,major\,\,axis}\,\,(a)$ & $0.3262\pm0.0059\,R_{\rm \odot}$  \\
   \hline
   \hline
   $\mathrm{Accretor\,\,mass}\,\,(M_{\mathrm WD})$ & $1.02\pm0.04\,M_{\rm \odot}$  \\
   \hline
   $\mathrm{Donor\,\,mass}\,\,(M_{\mathrm{Donor}})$ & $0.24\pm0.03\,M_{\rm \odot}$  \\
   \hline
   $\mathrm{Donor\,\,radius}\,\,(R_{\mathrm{Donor}})$ & $0.086\pm0.003\,R_{\rm \odot}$  \\
   \hline
   $\mathrm{Donor\,\,temperature}\,\,(T_{\mathrm{Donor}})$ & $9350\pm140\,\rm K$  \\
   \hline
   \hline

\end{tabular}
\caption{Table of parameters (the first five parameters are the astrometric solution reported by \emph{Gaia} eDR3\cite{Gaia2023}, at Epoch J2016.0 and Equinox J2000.0). The measured projected rotational velocity inferred from line broadening was not part of the broader joint parameter analysis, and is an independent measurement in excellent agreement with the value predicted by the joint analysis (predicted value: $226\pm8\,\mathrm{km\,s^{-1}}$).}
\end{table}

\newpage

\newpage

 \begin{addendum}
 \item K.B.B. is a former Pappalardo Postdoctoral Fellow in Physics at MIT and thanks the Pappalardo fellowship program for supporting his research.
 \item[Competing Interests] The authors declare that they have no
competing financial interests.
 \item[Correspondence] Correspondence and requests for materials
should be addressed to E.T.C.~(email: echickle@mit.edu).
\end{addendum}

\newpage 
\begin{methods}

\subsection{Period Search and Identification}

Period searching has a long history in astronomy, with established algorithms like Lomb-Scargle\citemethods{2018ApJS..236...16V} and Box Least Squares (BLS)\citemethods{2002A&A...391..369K}. Recent efforts by the Zwicky Transient Facility (ZTF) have systematically searched for periodic phenomena, discovering numerous ultracompact systems with orbital periods under approximately one hour [e.g., \citemethods{2020ApJ...905...32B}, \citemethods{2022MNRAS.512.5440V}]. However, searches for ultracompact binaries have been conducted almost exclusively in the Northern hemisphere, including ZTF which observes down to $\delta\approx30^\circ$ but suffers from sparse sampling at low declinations. To address this limitation and search the Southern hemisphere, we utilized data produced by two full-sky surveys, Asteroid Terrestrial-Impact Last Alert System (ATLAS; \citemethods{2018PASP..130f4505T}) and the Transiting Exoplanet Survey Satellite (TESS; \citemethods{2015JATIS...1a4003R}), to systematically identify light curves exhibiting periodic flux variations over short timescales. This search targeted a catalog of 1.3 million white dwarf candidates\citemethods{2021yCat..75083877G} using astronometric and photometric measurements from the Gaia Early Data Release 3 (EDR3) and validated with the spectroscopically confirmed white dwarf sample from the Sloan Digital Sky Survey (SDSS).

We search a grid of trial periods with equal spacing in frequency space. Our period range extends from 10 days down to the Nyquist limit of 400s for TESS light curves, and down to 2 minutes for ATLAS light curves, which have no Nyquist limit due to their irregular sampling. Given the multiple year baseline of ATLAS data, this necessitates searching $10^7$ frequencies for each of the more than one million sources. Our period search utilizes a GPU implementation of the Box Least Squares (BLS) algorithm\citemethods{2002A&A...391..369K}, using the \texttt{cuvarbase} package\footnote{\url{https://github.com/johnh2o2/cuvarbase}}. The analysis was performed on 16 Nvidia A100 GPUs distributed across four Linux nodes within a high-performance computing cluster.

We determine the best period by selecting the trial frequency that maximizes the periodogram power and select a subset of phase-folded light curves based on a combination of metrics to visually inspect. These metrics include periodogram peak significance, defined as the peak power compared to the standard deviation across the entire periodogram, and periodogram peak width, which helps mitigate spurious narrow peaks caused by aliasing and noise. We also consider the signal-to-noise ratio (SNR) of the eclipse, defined as the depth of the eclipse divided by the median absolute deviation. Additionally, we evaluate phase entropy, defined as the maximum difference in phase, to mitigate aliased signals. This multi-metric approach allows us to prioritize high-quality candidates for visual identification photometric behavior indicative of binarity, such as eclipses or asymmetric minima in ellipsoidal variables. This helps us distinguish true binary systems from pulsations, like those of Delta Scuti stars. We systematically conduct photometric and spectroscopic follow up on objects exhibiting compelling binary variability with an orbital period shorter than a hour. Follow-up of objects in this sample is ongoing.

\subsection{ATLAS Photometry} 

The four-telescope ATLAS system surveys the sky with a cadence of one day between $-50<\delta < +50$ and two days in the polar regions, using 0.5 m Wright-Schmidt telescopes. Two telescopes are located at the Haleakal\=a High Altitude Observatory and Mauna Loa Observatory in Hawaii, operational since 2015. The other two, located at El Sauce Observatory in Chile and Sutherland Observing Station in South Africa, became operational in early 2022, providing coverage in the Southern hemisphere. Each telescope has a 30 deg\textsuperscript{2} field of view, reaching a 5$\sigma$ limiting magnitude of approximately 19.7 in 30-second exposures in both the cyan (\emph{c}, covering 420-650 nm) and orange (\emph{o}, 560-820 nm) bands. In contrast, ZTF scans the Northern sky every two days with a 1.2 m telescope. For periodic phenomena, particularly those with short periods, a survey's sensitivity scales not just with aperture but also with the square root of the number of exposures, provided the systems are not fundamentally limited by read noise. Consequently, surveys with small telescopes but high sampling rates, such as ATLAS, can be comparable to large telescope surveys, like ZTF, for identifying short periodic phenomena.

ATLAS data, publicly available though its forced photometry website\footnote{\url{https://fallingstar-data.com/forcedphot/}}, provided 110 \emph{c}-band and 241 \emph{o}-band observations for ATLAS J1138-5139, spanning from January 7th, 2022 to January 26th, 2023. We combine data from multiple filters by computing the median magnitude in each filter and shifting \emph{o}-band so that its median magnitude matches \emph{c}-band data, in order to maximize the number of epochs. We remove data points with a zeropoint magnitude of greater than 17.5 and data points more than 3 IQR above the median or 10 IQR below the median to prevent clipping any data points from faint eclipses. We also convert timestamps to BJD prior to period searching.

\subsection{TESS Photometry} 

TESS is a space telescope in a 13.7-day orbit around the Earth that observes the sky in sectors measuring $24^\circ \times 96^\circ$ reaching a photometric precision of $\approx 10^{-2}$ in a 30-minute exposure at 16th TESS magnitude \citemethods{2015JATIS...1a4003R}.  During its second extended mission (Years 5 and 6, beginning September 2022), TESS published full-frame images (FFIs) at a 200 s cadence over a nearly continuous 27 days, with 3 to 4 hour pauses in data collection approximately every seven days for data downlinks, resulting in over ten thousand photometric measurements for each pointing.  Hence TESS' extraordinarily large field of view and high sampling rates makes it comparable to ZTF when it comes to short periodic phenomena. Most of the TESS footprint during Year 5 lies in the Southern ecliptic hemisphere, comprising of Sectors 61-69, whereas the scan of the Northern ecliptic hemisphere comprises of Sectors 56-60. ATLAS J1138-5139 was observed in Sector 64 Camera 2 CCD 1, providing 11,344 epochs. Existing TESS FFI light curve products prioritize publishing light curves for stars brighter than a TESS Magnitude of 16 and their open-source packages are intended for extraction of single or few targets as compared to the extraction of over a million light curves (e.g. the Quick-Look Pipeline (QLP; \citemethods{2020RNAAS...4..204H}), \texttt{eleanor} (\citemethods{2019PASP..131i4502F}), TESS-Gaia Light Curves (\texttt{tglc}; \citemethods{2023AJ....165...71H})). We performed forced aperture photometry on raw TESS FFIs at the coordinates of the Gaia EDR3 white dwarf candidates. The aperture radius and background annulus radii were tuned on known ultracompact systems, such as Gaia14aae. We detrend over 0.1-day windows to remove systematics associated with momentum dumps and scattered light from the Earth and Moon. We use QLP \citemethods{2020RNAAS...4..204H} quality flags to remove data points affected by, for example, cosmic rays and unstable pointing. The 3 sigma limiting magnitude is roughly 17 or 18 depending on how crowded the field is (not accounting for the effects of confusion noise, which can further dilute the signal). Hence an eclipse from an 18th magnitude source could not produce a more than 3 sigma outlier, and we are free to clip beyond that without fear of clipping away in-eclipse data points. We period search down to the Nyquist limit of 400 s and up to half the baseline (13.7 days). We do not detect signals from all objects in the target catalog, as some were too faint, had periods longer than the baseline, or suffered from blending due to 21 arcsecond pixels, especially in the Galactic plane.

\subsection{Magellan MagE observation}

On UT 2023 December 18 and 19, we obtained 4.5 hrs of phased-resolved spectroscopy using the Magellan Echellette (MagE) spectrograph mounted on the 6.5 m Magellan Baade telescope at Las Campanas Observatory. We utilized the 0.85" slit width, which was chosen to match the typical seeing conditions at Las Campanas and to balance the need for high spectral resolution with adequate light throughput, providing wavelength coverage of 3400 to 9400 \r{A} with a resolving power of $R\approx 4800$. The observations were made using the fast read speed mode of the MagE detector, reducing the dead time with minimal increase in read noise compared to the slow read speed. We employed 2x2 binning to improve the SNR by reducing the readout noise and increasing the effective signal per pixel. To minimize Doppler smearing of spectral lines, we opted for exposure times of 180 seconds. This duration corresponds to approximately 10\% of the orbital period of our target, ensuring Doppler smearing minimally broadens the spectral features, which are crucial for accurate radial velocity measurements. To ensure precise wavelength calibration we took Thorium-Argon (ThAr) arc lamp exposures at the telescope position of the object immediately following science exposures. The ThAr lamp provides a rich spectrum of 500 emission lines that serve as reference points for wavelength calibration. By taking the arc lamp exposures at the same telescope position, we account for any potential flexure or shifts in the instrument setup that could affect the wavelength solution. 

We reduced our data using the Pypeit data reduction pipeline\citemethods{pypeit:joss_pub}\citemethods{pypeit:zenodo}. For flux calibration, we use a standard star observed on a different night under similar conditions. The standard star observations were made with the same slit width, but observed with the Turbo readout speed and 1x1 binning. We manually bin the standard star data to match the science data binning, but do not correct the sensitivity function to account for the differences in readout noise and gain between the standard star and science observations.

\subsection{ULTRACAM observation}

We obtained high-speed photometric follow-up using the triple-beam CCD camera ULTRACAM\citemethods{2007MNRAS.378..825D} mounted on the 3.5 m New Technology Telescope (NTT) at the La Silla Observatory in Chile. We conducted a campaign of observations spanning serveral nights over the course of a year. For the observations, we used the Super SDSS $u_s$ as the blue channel filter, the Super SDSS $g_s$ as the green channel filter, and the Super SDSS $r_s$ as the red channel filter. ULTRACAM consists of frame-transfer chips, which takes data in the exposed area whilst data in the masked area is simultaneously being read out, effectively eliminating readout time overheads, allowing us to obtain as short as 3 s exposures. We used a combination of 3 and 6 s exposures in $g_s$ and $r_s$ filters and 3, 12, 18 s exposures in $u_s$ filters due to variable conditions across our nights of observing. For further details, please see Extended Data Table \ref{tab:Observations}. We reduced the ULTRACAM data using a publicly available pipeline\footnote{\url{https://cygnus.astro.warwick.ac.uk/phsaap/hipercam/docs/html/}}, masking nearby contaminating stars from the circular annulus centered on the target and using a dark frame from 2021. We perform aperture photometry with a variable radius, scaled to a multiple of the FWHM of the stellar profiles of each frame, with a smaller scale factor range chosen for the $u_s$ filter. The same extraction aperture is used for comparison stars, which are chosen to have a \emph{Gaia} BP/RP low-resolution spectra in order to perform synthetic photometry. The phase-folded and binned light curves from these observations can be seen in Figure \ref{fig:LC}, which exclude unstable observing conditions at the beginning and end of some observing runs. These light curves served as the basis for our analysis of the ellipsoidal modulation and eclipses exhibited by the luminous secondary and accretion disk, and as timing epochs in order to measure the orbital decay.

\subsection{Dark Energy Camera (DECam) photometry}
The Dark Energy Camera Plane Survey (DECaPS; \citemethods{2023ApJS..264...28S}) is an optical survey in the \emph{grizY} bands with the Dark Energy Camera (DECam; \citemethods{2015AJ....150..150F}) on the 4m Blanco telescope at the Cerro Tololo Inter-American Observatory(CITO). The DECaPS footprint covers the Galactic plane accessible in the southern hemisphere with $\delta\leq-24^\circ$ and contains 3.32 billion sources. We use DECaPS high-quality photometry to constrain the SED of J1138, providing a strong constraint on the temperature of the donor in the system.

\subsection{Swift observations}

We targeted ATLAS J1138-5139 with both the Ultra-Violet Optical Telescope (UVOT) and the X-Ray Telescope on the Neil Gehrels Swift Observatory, accumulating a total exposure time of 3 ks (ObsID: 00016298001, 00016298002). For the UVOT observations, we utilized the UVM2 filter, which is centered at 2246 \r{A}. The UVM2 filter was chosen because it has negligible red leak compared to the UVW1 and UVW2 filters, which allow an appreciable amount of light through at wavelengths greater than 300 nm. Hence, the UVM2 filter provides a more accurate constraint on the ultraviolet part of the spectral energy distribution (SED), which is used to constrain the donor WD properties. We conducted the UVOT observations in the Event mode, which records the arrival time of each photon. The source magnitude was derived from the UVOT image using the FTOOLS package \footnote{\url{http://heasarc.gsfc.nasa.gov/ftools}}\citemethods{2014ascl.soft08004N}.

In addition to the UVOT data, we obtained deeper XRT observations to probe the X-ray emission from ATLAS J1138-5139. However, the XRT observations resulted in a non-detection. Using the Living Swift XRT Point Source Catalogue\footnote{\url{https://www.swift.ac.uk/LSXPS/}}, we obtain a 3-$\sigma$ upper limit of $3.1\times 10^{-3}$ counts s$^{-1}$ corresponding to an upper limit on the unabsorbed flux of approximately $5.4\times 10^{30}$ erg s$^{-1}$ in the 0.2-8.0 keV bandpass. This upper limit was derived using the WebPIMMS Count Rate Simulator\footnote{\url{https://heasarc.gsfc.nasa.gov/cgi-bin/Tools/w3pimms/w3pimms.pl}} assuming a power law index of $\gamma=1.33$. The assumed power-law index is based on the spectral characteristics observed in a similar system, ZTF J0127+5258, an edge-on white dwarf binary with an accretion disk in a 13.7-minute orbital period, studied with Chandra ACIS-I in a 16 ks observation\citemethods{2023ApJ...953L...1B}.

\subsection{Radial velocity analysis}

To measure the radial velocities of ATLAS J1138-5139, we utilized the high-resolution MagE spectroscopic data to analyze the Balmer series of absorption lines as well as the narrow Ca II K absorption line. We measured the velocities by simultaneously fitting Voigt profiles to the Balmer absorption lines within a single exposure. The fitting process involved minimizing the difference between the observed line profiles and the model Voigt profiles using a least-squares optimization technique. From the simultaneous fit, we obtained a single radial velocity measurement for each exposure. The uncertainties in the radial velocity measurements were estimated from the covariance matrix of the fit.

We then fit a sinusoidal radial velocity curve to all of the epochal RV measurements and minimize the $\chi^2$ error of this fit. We accounted for the small degree of Doppler smearing in this analysis.

\subsection{Rotational broadening analysis} 
In addition to using the MagE spectra to estimate the radial velocity semi-amplitude, we also used the narrow Ca II K absorption line at $3933\rm\,\AA$ to estimate the rotational Doppler broadening of the line. We performed the fit for this by using the pyastronomy line broadening modules, first broadening a metal polluted WD atmospheric model using the instrumental broadening of 4800, and then further broadening this with rotational broadening, which yielded our estimate of $237.3\pm12.5\,\text{km}\,\text{s}^{-1}$. This is in excellent agreement with the value predicted by the joint analysis used to infer the component masses and the donor radius in the binary (we did not include the rotational broadening measurement as a constraint in the joint analysis, but instead chose to use the measurement as an independent verification of the robustness of our analysis.

\subsection{Dust extinction analysis}

Accurate modeling of the spectral energy distribution (SED) requires a careful assessment of interstellar dust extinction. To estimate the effect of dust on the system's photometry, we utilized a 3D map of interstellar dust extinction\citemethods{2024A&A...685A..82E}. This map provides the spatial distribution of dust out to a distance of 1.25 kpc from the Sun with parsec-scale resolution. The map indicates negligible reddening along the line of sight to ATLAS J1138-5139, with an estimated $E(B-V)\approx 10^{-4}$, resulting from its location above the Galactic mid-plane $b\approx 9.6^\circ$, as seen in Extended Data Figure \ref{fig:dust}.

\begin{figure}
\includegraphics[width=\textwidth]{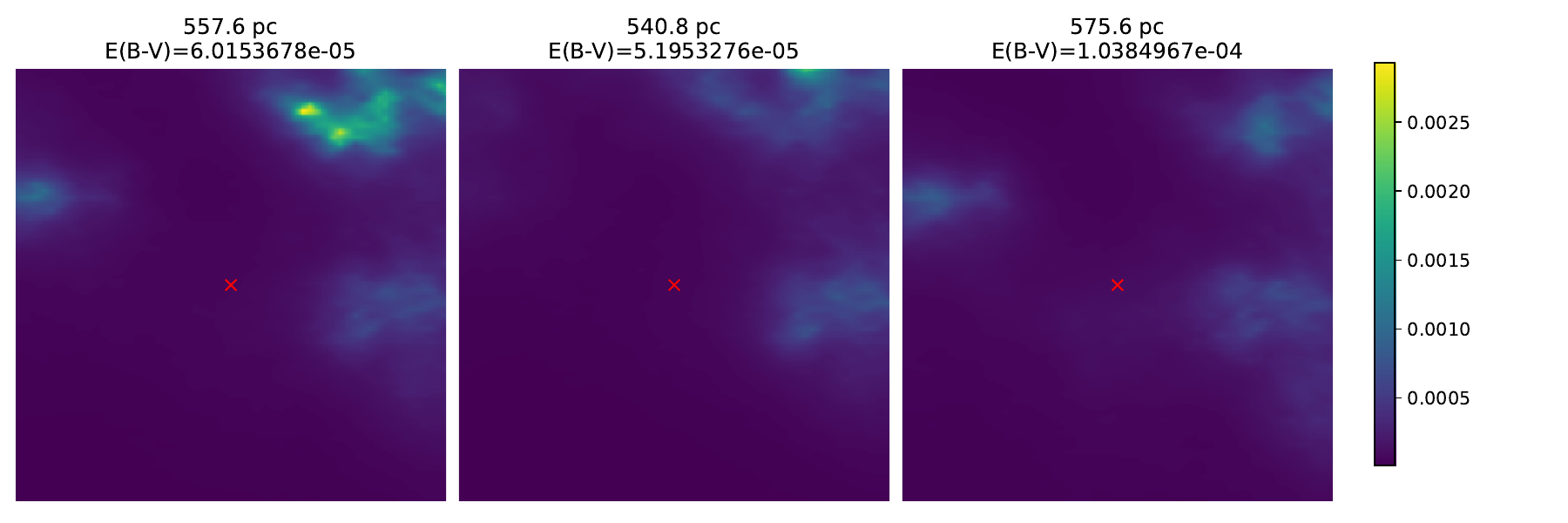}
\linespread{1.3}\selectfont{}
\caption{Interstellar dust extinction maps illustrating the extinction along the line of sight to ATLAS J1138-5139 at three distances. The first column corresponds to the system's observed distance $d=557$ pc, the second column shows the extinction at $d-\sigma_{d,\text{lower}}=541$ pc, and the third column represents $d+\sigma_{d,\text{upper}}=576$ pc, where $\sigma_d$ is the uncertainty in the distance measurement. The location of the system is marked with a red ``X" in all panels. The values of $E(B-V)$ are negligible across all distances.}
\label{fig:dust}
\end{figure}

\subsection{Joint analysis and parameter estimation}

We perform a joint analysis that considers our astrometric, spectroscopic, and photometric constraints simultaneously to determine the physical parameters of ATLAS J1138-5139. The free parameters in our model are the donor temperature, $\cos{i}$ (where $i$ is the inclination angle), the distance, and the component masses. 

The distance to ATLAS J1138-5139 was constrained using \emph{Gaia} astrometric measurements. We included this information in the joint analysis to strongly constrain the distance estimate.

Given the observed eclipses in the lightcurve, we set a lower bound on the inclination of 76 degrees given the mass ratio in the system. To appropriately weight this in our joint analysis, we took a uniform distribution in cos($i$) from 0 to 1, and truncated it to range from 0 to 0.24 to reflect the lower bound of 76 degrees on the inclination implied by the presence of eclipses.

The photometric constraint in the likelihood function is derived from from comparing synthetic photometry of Extremely Low Mass (ELM) white dwarf atmosphere models\citemethods{2009ApJ...696.1755T,2011ApJ...730..128T} to the observed spectral energy distribution (SED), which is dominated by the flux output of the lower mass white dwarf. These models provide the theoretical flux distribution for white dwarfs with different temperatures and surface gravities. We scale the model atmospheres to physical units of erg s$^{-1}$ cm$^{-2}$ \r{A}$^{-1}$ using the distance (which is a free parameter) and by fixing the donor radius as consistent with  Roche-filling, effectively fixing the density of the donor \citemethods{1983ApJ...268..368E}. The mean density of Roche lobe-filling stars at a given orbital period can be conveniently approximated:
\[P(\rho)^{1/2} = 0.1375\left(\frac{q}{(1+q)}\right)^{1/2}r_L^{-3/2}\]
where $r_L$ is in units of orbital separation $a$, which we evaluate using Kepler's law:
\[ \Omega^2 = \left(\frac{2\pi}{P_{\text{orb}}}\right)^2 = \frac{GM}{a^3}\]
\[a = \left(GM \left(\frac{P_{\text{orb}}}{2\pi}\right)^2\right)^{1/3} \]
We calculate the likelihood by comparing the synthetic photometric measurements from the WD atmosphere model to the observed SED. Our SED included the following measurements: Swift UVOT, ULTRACAM, and DECaPS photometry.

The spectroscopic constraint in the likelihood function is derived from the radial velocity semi-amplitude observed with the MagE data. In the likelihood function of the joint anlaysis, the radial velocity is included in the binary mass function:
\[\begin{split}
    \frac{(M_1\sin i)^3}{(M_1+M_2)^2} &= \frac{P_{\text{orb}}K_2^3}{2\pi G} 
\end{split}\]

The joint analysis was performed using a Kernel Density Estimate (KDE)-based approach to explore the parameter space efficiently, by utilizing the library \texttt{UltraNest}\footnote{\url{https://johannesbuchner.github.io/UltraNest/readme.html}}. The free parameters (donor temperature, $\cos i$, distance, and component masses) were varied simultaneously to find the best-fit values that match the observed data. The derived parameters are reported in Table 1 in the main text.

\subsection{Light curve analysis}

While we chose not to include a lightcurve model in constraining the parameters reported for the system, we nonetheless constructed a toy model to demonstrate that accounting for accretion was needed to describe the morphology of the lightcurve. The reason we chose to omit lightcurve modeling from our parameter estimation is because it requires a large number of degrees of freedom (due to the presence of a disk and hot spot), whereas our parameter estimates come from a much more simple and robust set of constraints. We used the \texttt{LCURVE} code\citemethods{2010MNRAS.402.1824C} to model the ULTRACAM \emph{u}-band, \emph{g}-band, and \emph{r}-band light curve of ATLAS J1138-5139. The free parameters in our light curve included: the donor temperature (\texttt{t2}), inclination angle, exponent of surface brightness over disk (\texttt{texp\_disc}), the length scale of bright spot (\texttt{length\_spot}), the surface brightness of the spot (\texttt{temp\_spot}), and the fraction of the spot taken to be equally visible at all phases (\texttt{cfrac\_spot}). We fix the mass ratio (\texttt{q}), the accretor radius (\texttt{r1}) and donor radius (\texttt{r2}). We fix the accretor radius using the mass-radius relation for white dwarfs \citemethods{1961ApJ...134..683H}. We obtained gravity-darkening and limb-darkening coefficients from \citemethods{2020yCat..36340093C}.

The disk and hot spot were crucial for accurately reproducing the observed variations in brightness, successfully modeling key features of the light curve, including the shape of the eclipses. A purely detached binary model failed to account for the observed light curve morphology. A best fit example model is shown in Extended Data Figure \ref{fig:LCURVE}.

\begin{figure}
\includegraphics[width=6.5in]{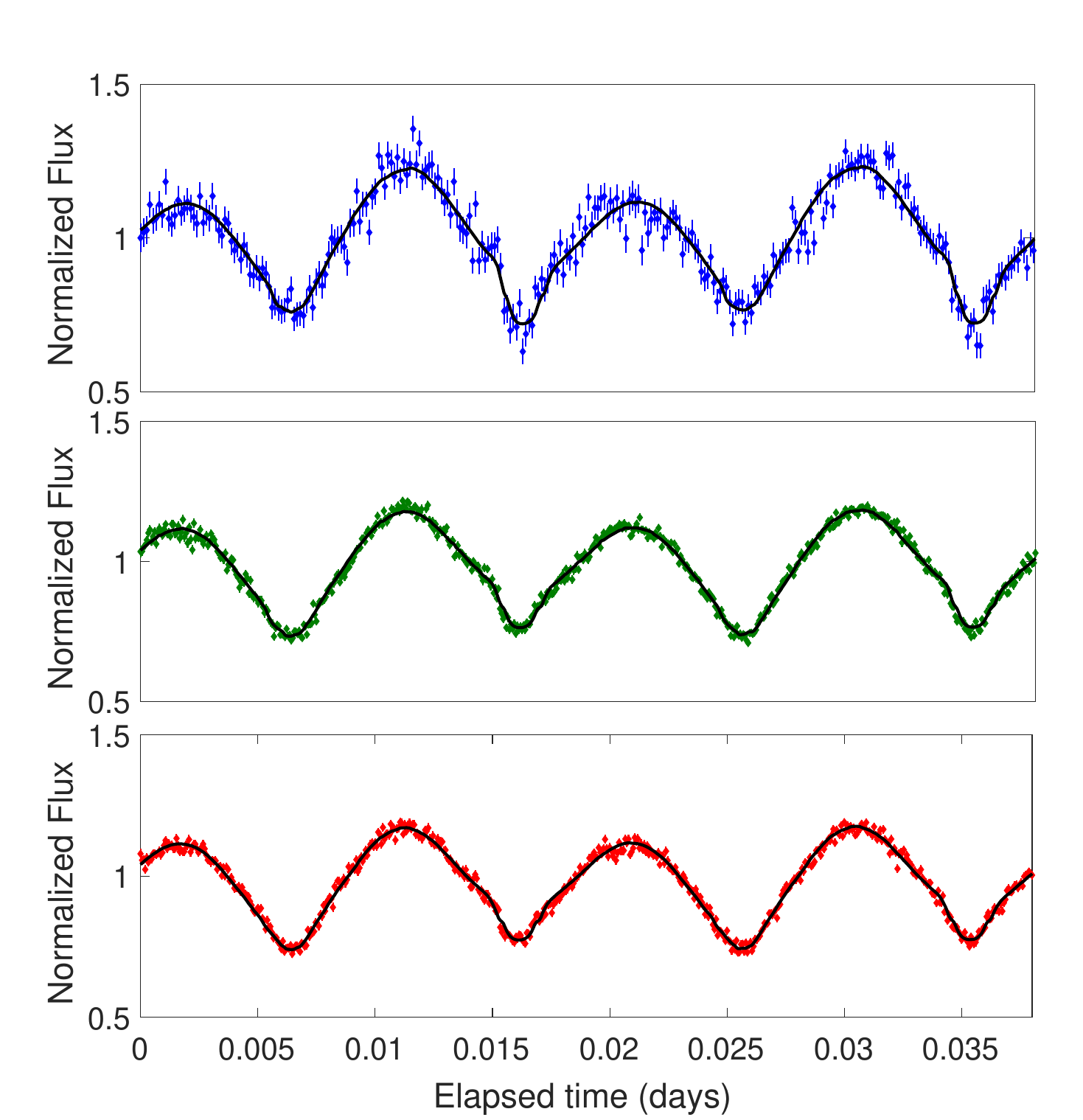}
\linespread{1.3}\selectfont{}
\caption{Three color ULTRACAM light curve from March 20, 2023, overlaid with the best-fit toy \texttt{LCURVE} model (solid black lines). The model includes contributions from an accretion disk and a hot spot where the accretion stream intersects the outer disk, which were needed to achieve an acceptable fit to the lightcurve. The best fit model gives a near edge-on inclination (88.6 degrees), though we chose not to include this model as part of our joint fit, due to the large number of model-dependent free parameters involved, as compared to our more simple and robust approach.}
\label{fig:LCURVE}
\end{figure}

\subsection{Kinematic Analysis}

We conducted a kinematic analysis of ATLAS J1138-5139's Galactic orbit (see Extended Data Figure \ref{fig:kinematic}) and found it to be consistent with residing at the boundary between the Galactic thin and thick disk\citemethods{2021AstL...47..534B}, orbiting between 1.2 and 2.7 kpc from the Galactic center. We used the galpy\citemethods{2015ApJS..216...29B} package to compute its trajectory around the Milky Way over 6 Gyr, using the MWPotential2014 potential\citemethods{2015ApJS..216...29B}.

\begin{figure}
\includegraphics[width=\textwidth]{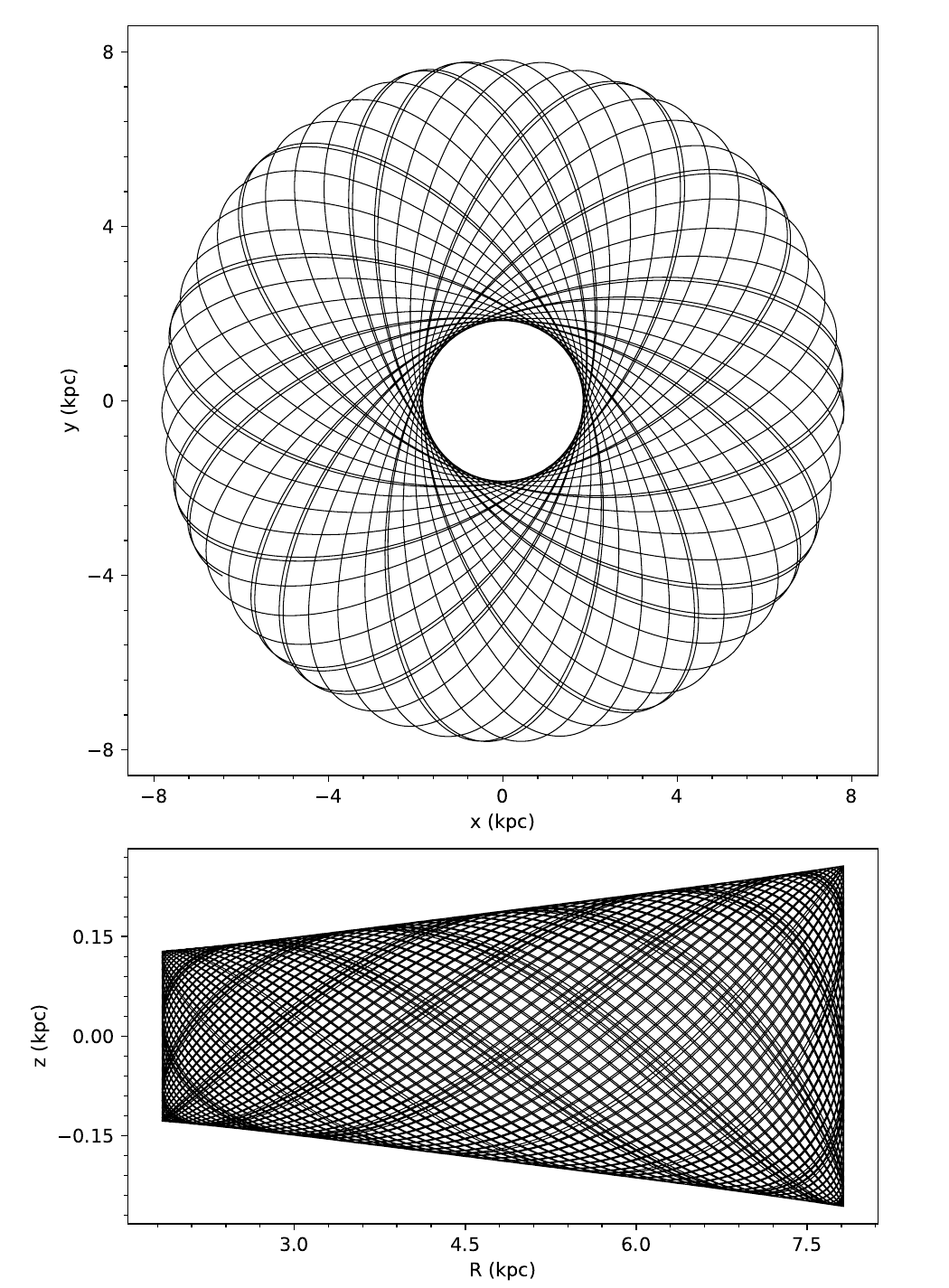}
\linespread{1.3}\selectfont{}
\caption{A set of panels depicting the orbit of ATLAS J1138-5139 around the Milky Way over the next 6 Gyr, at the boundary of the Galactic thin and thick disk populations.}
\label{fig:kinematic}
\end{figure}

\subsection{LISA Analysis} 


The GW signals of non-eccentric ultracompact binaries such as ATLAS J1138-5139 can be modeled with 8 parameters: the initial frequency ($f_0$) and time derivative of the frequency ($\dot{f}_0$), the amplitude ($\mathcal{A}$), two angles encoding the sky position ($\alpha$, $\beta$), and three angles encoding the binary's orbital angular momentum vector with respect to the observer, which are the inclination ($\iota$), polarization angle ($\psi$), and initial phase ($\phi_0$). Electromagnetic constraints on inclination and sky position ($\alpha$, $\beta$) can reduce the GW-only parameter uncertainties, including polarization angle ($\psi$), initial GW phase ($\phi$), and amplitude ($\mathcal{A}$), by up to a factor of two\citemethods{2012A&A...544A.153S,2013A&A...553A..82S}. The strain amplitude itself is a function of the chirp mass, the distance to the binary, and the frequency\citemethods{1989ApJ...345..434T}:
\[S= \frac{2(G\mathcal{M})^{5/3}(\pi f_{\text{GW}})^{2/3}}{c^4D}.  \]
The characteristic strain is then obtained from multiplying by the (square root of the) total number of orbital cycles observed over the total observational baseline ($T_{\text{obs}}$), $h_c = S\times \sqrt{f_{\text{GW}}T_{\text{obs}}}$ \citemethods{2015CQGra..32a5014M}. For a roughly monochromatic source, the signal-to-noise ratio (SNR) is obtained from the ratio of $h_c$ to the $LISA$ sensitivity at a given frequency\citemethods{2022ApJS..260...52W}.




The timescale over which the system will merge due to gravitational waves is given by:
\[\tau = 47925\frac{(M_1+M_2)^{1/3}}{M_1M_2}P^{8/3}\,\text{Myr}\]
where the masses are in units of $M_\odot$, the orbital period $P$ is in days, and $\tau$ is in Myr\citemethods{1962ApJ...136..312K}. For the case of ATLAS J1138-5139, using the masses derived from RV measurements results in $\tau=8$ Myr, i.e. the system will merge in significantly less than a Hubble time.

\end{methods}

\section{Data Availability}
The observational data used in this work is either publically available (e.g. ATLAS and TESS), or will be made available upon request to the corresponding author.
\section{Code Availability}
The python code used to perform the analyses in this work will be made available by the corresponding author upon request.


\newpage 



{\small
\renewcommand{\arraystretch}{1.2} 
\setlength{\tabcolsep}{4pt}      
\setcounter{table}{0}

\begin{longtable}{ccccc}
    \hline
    \hline
    Instrument & Filter & Date & \# of Exposures & Exposure Time \\
    \hline
    \endfirsthead
    \hline
    \hline
    Instrument & Filter & Date & \# of Exposures & Exposure Time \\
    \hline
    \endhead
    \hline
    \endfoot
    \hline
    \hline
    \endlastfoot
    
    ULTRACAM & Super u' & Mar 8 2023 & 649 & 12s \\
    ULTRACAM & Super g' & Mar 8 2023 & 649 & 6s \\
    ULTRACAM & Super r' & Mar 8 2023 & 649 & 6s \\ 
    \hline 
    ULTRACAM & Super u' & Mar 10 2023 & 469 & 12s \\
    ULTRACAM & Super g' & Mar 10 2023 & 469 & 6s \\
    ULTRACAM & Super r' & Mar 10 2023 & 469 & 6s \\    
    \hline 
    ULTRACAM & Super u' & Mar 19 2023 & 451 & 18s \\
    ULTRACAM & Super g' & Mar 19 2023 & 451 & 6s \\
    ULTRACAM & Super r' & Mar 19 2023 & 451 & 6s \\  
    \hline 
    ULTRACAM & Super u' & Feb 6 2024 & 1373 & 3s \\
    ULTRACAM & Super g' & Feb 6 2024 & 1373 & 3s \\
    ULTRACAM & Super r' & Feb 6 2024 & 1373 & 3s \\ 
    \hline
    ULTRACAM & Super u' & Feb 10 2024 & 1719 & 3s \\
    ULTRACAM & Super g' & Feb 10 2024 & 1719 & 3s \\
    ULTRACAM & Super r' & Feb 10 2024 & 1719 & 3s \\     
    \hline
    ULTRACAM & Super u' & Feb 11 2024 & 553 & 3s \\
    ULTRACAM & Super g' & Feb 11 2024 & 553 & 3s \\
    ULTRACAM & Super r' & Feb 11 2024 & 553 & 3s \\  
    \hline
    ULTRACAM & Super u' & July 7 2024 & 1516 & 3s \\
    ULTRACAM & Super g' & July 7 2024 & 1516 & 3s \\
    ULTRACAM & Super r' & July 7 2024 & 1516 & 3s \\      
    \hline 
    MagE & & Dec 18 2023 & 30 & 180s \\
    MagE & & Dec 19 2023 & 61 & 180s \\
    \hline 
    Swift XRT & & Oct 23 2023 & 1 & 1151.2478 \\
    Swift XRT & & Oct 27 2023 & 1 & 1866.83300 \\
    \hline 
    Swift UVOT & UVM2 & Oct 23 2023 & 1 & 1153.760 \\
    Swift UVOT & UVM2 & Oct 27 2023 & 1 & 1865.19200 \\
    \hline
\end{longtable}
\captionof{table}{Table of observations.} 
\label{tab:Observations}
}

\end{document}